\documentclass[nofootinbib]{revtex4}

\usepackage{graphicx}
\begin{document}
\bibliographystyle{revtex}

\preprint{P1WG3_kalinowski_0716}
\preprint{IFT-01-32}
\preprint{DESY??}
\title{SUSY Parameters from Charginos}



\author{Jan Kalinowski}
\email[]{Jan.Kalinowski@fuw.edu.pl}
\thanks{Preprint number: IFT-01-32 }
\affiliation{Instytut Fizyki Teoretycznej, Uniwersytet Warszawski,
Warsaw, Poland} 
\author{Gudrid Moortgat-Pick}
\email[]{gudrid@mail.desy.de}
\affiliation{Deutsches Electronen Synchrotron, Hamburg, Germany}



\begin{abstract}
The chargino pair production processes at $e^+e^-$
collisions are explored to reconstruct the fundamental SUSY
parameters: the SU(2) gaugino parameter $M_2$, the
higgsino mass parameter $\mu$ and $\tan\beta$. Both CP-conserving and
CP-violating SUSY sectors are discussed.
\end{abstract}

\maketitle

\section{Introduction}
The concept of symmetry between bosons and
fermions, supersymmetry (SUSY), 
has so many attractive features that the supersymmetric
extension of the Standard Model is still widely considered as a most natural
scenario. However, if realized in Nature, supersymmetry must be 
broken at low energy
since no superpartners of ordinary particles have been observed so
far.  Technically it is achieved  by introducing the
soft--supersymmetry breaking parameters: 
gaugino masses $M_i$, sfermion masses $m_{\tilde{f}}$ and trilinear
couplings $A^f$ (gauge group and generation indices are understood). 
This gives rise to
a large number of parameters. Even in the minimal supersymmetric model (MSSM) 
105 new parameters are introduced.  This number of
parameters, reflecting our ignorance of SUSY breaking mechanism, 
can be reduced by additional physical assumptions.

After discovering supersymmetric particles, however,  the priority
will be to determine the low-energy Lagrangian parameters. They 
should be measured independently of any theoretical
assumptions. This will
allow us to verify the relations among them, if any,  in order to distinguish
between various SUSY models.

Here we  outline how the fundamental SUSY parameters:
the SU(2) gaugino parameter $M_2$, the
higgsino mass parameter $\mu$ and $\tan\beta$, 
can be determined from the
measurements of chargino pair production cross sections with polarized
beams at future $e^+e^-$ linear colliders. The results
summarized here have been worked out in a series of papers
\cite{choi1}, to which we refer for  
more detailed discussions  and references.

\section{Chargino sector}
After the electroweak symmetry breaking, the mass matrix of the 
spin-1/2 partners of the
$W^{\pm}$ gauge bosons and the charged Higgs bosons, $\tilde W^\pm$
and $\tilde H^\pm$, is non diagonal
\begin{eqnarray}
{\cal M}_C=\left(\begin{array}{cc}
                M_2                &      \sqrt{2}m_W\cos\beta  \\
             \sqrt{2}m_W\sin\beta  &             \mu   
                  \end{array}\right)\
\label{eq:mass matrix}
\end{eqnarray}
The mass eigenstates, the two charginos $\tilde{\chi}^\pm_{1,2}$,  
are mixtures
of the charged SU(2) gauginos and higgsinos. 
Since the chargino mass matrix ${\cal M}_C$ is not symmetric, two
different 
unitary matrices acting on the left-- and right--chiral 
$(\tilde{W},\tilde{H})_{L,R}$ two--component 
states 
\begin{eqnarray}
U_{L,R}\left(\begin{array}{c}
             \tilde{W}^- \\
             \tilde{H}^-
             \end{array}\right)_{L,R} =
       \left(\begin{array}{c}
             \tilde{\chi}^-_1 \\
             \tilde{\chi}^-_2
             \end{array}\right)_{L,R} 
\end{eqnarray}
are needed to diagonalize the matrix eq.(\ref{eq:mass matrix}).
In general CP-noninvariant theories 
the mass parameters are complex. 
However, by reparametrization of the fields, 
$M_2$ can be assumed real and positive without loss of generality so that
the only non--trivial reparametrization--invariant phase 
may be attributed to 
$\mu=|\mu|\,{\rm e}^{i\Phi_\mu}$ with $0 \leq \Phi_\mu \leq 2\pi$.
The unitary matrices $U_L$ and $U_R$ can  be parameterized in the
following way:
\begin{eqnarray}
 U_L=\left(\begin{array}{cc}
             \cos\phi_L & {\rm e}^{-i\beta_L}\sin\phi_L \\
            -{\rm e}^{i\beta_L}\sin\phi_L & \cos\phi_L
             \end{array}\right), \qquad
 U_R=\left(\begin{array}{cc}
             {\rm e}^{i\gamma_1} & 0 \\
             0 & {\rm e}^{i\gamma_2}
             \end{array}\right)
        \left(\begin{array}{cc}
             \cos\phi_R & {\rm e}^{-i\beta_R}\sin\phi_R \\
            -{\rm e}^{i\beta_R}\sin\phi_R & \cos\phi_R
             \end{array}\right) 
\end{eqnarray}
The mass eigenvalues $m^2_{\tilde{\chi}^\pm_{1,2}}$ are given by
\begin{eqnarray}
m^2_{\tilde{\chi}^\pm_{1,2}}
   =\frac{1}{2}\left[M^2_2+|\mu|^2+2m^2_W\mp \Delta_C\right]
\end{eqnarray}
with $\Delta_C$ involving the phase  $\Phi_\mu$: 
\begin{eqnarray}
\Delta_C=[(M^2_2-|\mu|^2)^2+4m^4_W\cos^2 2\beta
              +4m^2_W(M^2_2+|\mu|^2)+8m^2_WM_2|\mu|
               \sin2\beta\cos\Phi_\mu]^{1/2}
\end{eqnarray}
%
The four phase angles $\{\beta_L,\beta_R,\gamma_1,\gamma_2\}$
are not independent: they are functions  of  the invariant angle
$\Phi_\mu$ and their explicit form can be found in \cite{choi1}. 
All four phase angles vanish in CP--invariant theories for which 
$\Phi_\mu = 0$ or $\pi$.  The rotation angles 
$\phi_L$ and $\phi_R$ satisfy the relations:
\begin{eqnarray}
&&c_{2L,R}\equiv \cos 2\phi_{L,R}
=-\left[M_2^2-|\mu|^2\mp 2m^2_W\cos 2\beta\right]/\Delta_C
   \nonumber\\ 
&&s_{2L,R}\equiv \sin 2\phi_{L,R}
=-2m_W\, [M^2_2+|\mu|^2\pm(M^2_2-|\mu|^2)\cos 2\beta
                     +2M_2|\mu|\sin2\beta\cos\Phi_\mu]^{1/2}/\Delta_C
\end{eqnarray}
%
\section{Inverting}
From the set $m_{\tilde{\chi}^\pm_{1,2}}$ and $\cos 2\phi_{L,R}$  
the fundamental supersymmetric parameters 
$\{M_2, |\mu|, \cos\Phi_\mu, \tan\beta\}$ in CP--(non)invariant theories 
can be determined unambiguously in the following way:
\begin{eqnarray}
&&  M_2=m_W[\Sigma-\Delta(c_{2L}+c_{2R})]^{1/2}\nonumber\\ 
&&  |\mu|=m_W[\Sigma+\Delta (c_{2L}+c_{2R})]^{1/2}\nonumber\\
&&\cos\Phi_\mu=[\Delta^2 (2-c^2_{2L}-c^2_{2R})-\Sigma]
                   \{[1-\Delta^2 (c_{2L}-c_{2R})^2]
                         [\Sigma^2 -\Delta^2(c_{2L}+c_{2R})^2]\}^{-1/2}
\nonumber \\
&&\tan\beta =\{[1-\Delta (c_{2L}-c_{2R})]/[
                      1+\Delta (c_{2L}-c_{2R})]\}^{1/2}
\label{inv}
\end{eqnarray}
where we introduced the abbreviations
$ \Sigma
=(m^2_{\tilde{\chi}^\pm_2}+m^2_{\tilde{\chi}^\pm_1}-2m^2_W)/2m^2_W$,
and 
$ \Delta=(m^2_{\tilde{\chi}^\pm_2}-m^2_{\tilde{\chi}^\pm_1})/4m^2_W$. 
Therefore to reconstruct the above parameters the chargino masses and
$\cos2\phi_{L,R}$ have to be measured independently. This can be done
from the measurements of  the production of chargino pairs at $e^+e^-$
colliders, where they  are produced at tree level via  
$s$--channel $\gamma$ 
and $Z$ exchanges, and $t$--channel $\tilde{\nu}_e$ exchange.   

The chargino masses can be measured  very precisely  from the sharp
rise of the cross sections at threshold \cite{LCWS}.  The mixing angles
$\phi_{L,R}$ on the other hand can be determined from measured 
cross sections for the chargino production with polarized beams.  
For this purpose polarized beams are crucial since the mixing angles
$\phi_{L,R}$ encode the chiral dependence of the chargino
couplings to the $Z$ gauge boson and to the electron-sneutrino
current. All the production cross sections 
$\sigma_\alpha \{ij\}=\sigma(e^+e^-_{\alpha} \rightarrow
\tilde{\chi}^+_i\tilde{\chi}^-_j)$ for any  
beam polarization $\alpha$ and for any 
combination of chargino pairs $\{ij\}$
 depend only on  
$\cos 2 \phi_L$
and $\cos 2 \phi_R$ apart from the chargino masses, the sneutrino
mass and the Yukawa $\tilde{W} e \tilde{\nu}$ coupling.\footnote{The
explicit dependence on the $\sin 2 \phi_{L,R}$ and on 
the phase angles $\beta_{L},\, \beta_{R},\, \gamma_1,\,
\gamma_2$ has to disappear from CP-invariant quantities. 
The final ambiguity in $\Phi_\mu \leftrightarrow 2 \pi - \Phi_\mu$ 
in CP--noninvariant theories must be resolved by measuring observables
related to the 
normal $\tilde{\chi}^-_1$ or/and $\tilde{\chi}^+_2$  polarization in 
non--diagonal $\tilde{\chi}^-_1\tilde{\chi}^+_2$ chargino--pair 
production.}
In fact the cross sections are binomials 
in the [$\cos2\phi_L,\cos2\phi_R$] plane. Therefore any two contour
lines, $\sigma_L\{11\}$ and $\sigma_R\{11\}$ for example, 
will at least cross at one point in the plane between $-1 \leq \cos2\phi_L,
\cos2\phi_R \leq +1$.  However, the contours, being ellipses or hyperbolae, 
may cross 
up to four times. Imposing contours
for  other cross sections $\sigma_\alpha\{ij\}$ this ambiguity can be
resolved and,  
at the same time, the sneutrino mass and the identity between the
$\tilde W e \tilde\nu $
Yukawa  and the $W e \nu$ gauge couplings can be tested.
The sneutrino exchange does not contribute for the right-handed
polarized electron beams, $\alpha=R$. Therefore, while the curves for
$\sigma_R\{ij\} $ are fixed, the curves
for $\sigma_L\{ij\}$ will move in the $[\cos2\phi_L,\cos2\phi_R]$ plane with 
changing $m_{\tilde\nu}$ and the Yukawa coupling. All curves will
intersect  in the same point only if the mixing angles as well as the
sneutrino  
mass and the Yukawa coupling correspond to the correct physical values. 

It has been checked that combining the analyses of 
$\sigma_{R}\{ij\}$ and $\sigma_{L}\{ij\}$, the masses, the mixing
parameters and the Yukawa coupling can be determined to quite a high 
precision. For example, for the reference point RR1 
introduced in Ref.\cite{LCWS}, defined by 
$M_{2}=152$ GeV, $\mu=316$ GeV and $\tan\beta=3$,  one
can expect %
\begin{eqnarray}
&&{ }\hskip -1.2cm m_{\tilde\chi_1^\pm}=128\pm 0.04\, {\rm GeV}\quad 
   \cos2\phi_L=0.645 \pm 0.02  \qquad
   g_{\tilde{W}e\tilde{\nu}}/g_{We\nu}= 1\pm 0.001 \nonumber\\  
&&{ }\hskip -1.2cm m_{\tilde\chi_2^\pm}=346\pm 0.25\, {\rm GeV}\quad
   \cos2\phi_R=0.844 \pm 0.005 
\label{eq:measured}
\end{eqnarray}
where the 1$\sigma$ statistical errors are for an integrated 
luminosity of $\int {\cal L} =1\,\,{\rm ab}^{-1}$ collected at $\sqrt{s}=800$
GeV.

Using the eqs.(\ref{inv}) 
 the accuracy which can be expected in such an analysis 
for two CP--invariant reference points, the  RR1 defined above and the 
RR2 defined by $M_{2}=150$ GeV, $\mu=263$ GeV and $\tan\beta=30$, is
as follows (errors are 1$\sigma$ statistical only assuming 100\%
polarized beams)
\begin{center}
\begin{tabular}{ccccc}
$M_2$      & ~~~~~~~~~& $152\pm 1.75$\, GeV &~~~~~~~~~  
&$150\pm  1.2$\, GeV \\ 
$\mu$      && $316\pm 0.87$\, GeV && $263\pm 0.7$\, GeV \\ 
$\tan\beta$&&  $3\pm 0.69$         & & $> 20.2$ 
\end{tabular}
\end{center}
where the first (second) column is for RR1 (RR2). 
If $\tan\beta$ is large, this parameter is difficult to
extract from the chargino sector. Since the chargino observables depend 
only on $\cos2\beta$, the dependence on $\beta$ is flat for 
$2\beta\rightarrow \pi$ so that eq.(\ref{inv}) is not very 
useful to derive the value of
$\tan\beta$ due to error propagation. A significant lower bound can
be derived nevertheless in any case. 

The errors derived above have been obtained assuming that the
sneutrino mass is known from {\it e.g.} sneutrino pair production.  If the
sneutrinos, however, are beyond the kinematical reach, their masses
can be inferred from the
forward--backward asymmetries of the decay leptons \cite{MPF}. 
For high precision experimental analyses also radiative
corrections should be included \cite{rc}. 
\section{Incomplete chargino system}
For the above analyses the knowledge of both chargino masses is
crucial. However, at  
an early phase of the $e^+e^-$ linear collider the energy may only be
sufficient to reach the threshold of the light chargino pair
$\tilde{\chi}^+_1\tilde{\chi}^-_1$. Nearly the entire structure of the 
chargino system can nevertheless be reconstructed even in this case.

From the $\sigma_L\{11\}$ and $\sigma_R\{11\}$ the mixing angles
$\cos2\phi_L$ and $\cos2\phi_R$ can be determined up to at most a
four--fold ambiguity assuming that the sneutrino mass and the Yukawa
coupling are known.  The ambiguity can be resolved within the chargino
system by adding the information from measurements with transverse
beam polarization or by analyzing the polarization of the charginos in
the final state and their spin--spin correlations \cite{choi1}.  
The knowledge of
$\cos2\phi_L$, $\cos2\phi_R$ and $m_{\tilde\chi_1^\pm}$ is sufficient
to derive the fundamental gaugino parameters $\{M_2,\mu,\tan\beta\}$
in CP--invariant theories up to at most a discrete two--fold
ambiguity.  This remaining ambiguity can be removed by {\rm e.g.}
confronting the ensuing Higgs boson mass $m_{h^0}$ with the
experimental value. Alternatively, the ambiguity can also be resolved
by analyzing the light neutralino $\tilde{\chi}^0_1\,
\tilde{\chi}^0_2$ system for left and right polarized beams. At the
same time the U(1) gaugino mass parameter $M_1$ can also be determined
\cite{ckmz}.

\section{Conclusions}
The measured
chargino masses $m_{\tilde{\chi}^\pm_{1,2}}$ and the 
two mixing angles $\phi_L$ and $\phi_R$ are enough to extract the fundamental 
SUSY parameters $\{M_2,|\mu|, \cos\Phi_{\mu}, \tan\beta\}$ 
unambiguously;  a discrete two--fold ambiguity $\Phi_\mu \leftrightarrow
2\pi-\Phi_\mu$  can be
resolved only by measuring the CP-violating observable.

%
%

%
%

\begin{acknowledgments}
JK was supported in part by the KBN Grant 5 P03B 119 20 (2001-2002) and 
the European Commision 5-th framework contract HPRN-CT-2000-00149. 
GMP was partially supported by the DPF/Snowmass Travel Fellowship from the
DPF of the American
Physical Society, and of the Snowmass 2001 Organizing Committee.
We thank S.Y. Choi and P.M. Zerwas for many discussions and fruitful
collaboration. 
\end{acknowledgments}


\end{document}